\documentclass[12pt]{article}
\usepackage{epsfig,amssymb,amsmath}
\usepackage{jheppub}

\usepackage{esint} 
\usepackage{breqn}

\usepackage{soul}

\def \tr {\mathop{\rm tr}\nolimits}

\def \e  {\mathop{\rm e}\nolimits}

\newcommand\lr[1]{{\left({#1}\right)}}

\newcommand \vev [1] {\langle{#1}\rangle}

\newcommand\re[1]{(\ref{#1})}
\def \qqquad {\qquad\quad}
\def \qqqquad {\qquad\qquad}

\def\numberbysection{\@addtoreset{equation}{section}
                     \def\theequation{\thesection.\arabic{equation}}}

\preprint{\small  \parbox[t]{25mm}{IPhT-T16/036}}

\title{\Large Revisiting instanton corrections to the Konishi multiplet}

\author[a]{Luis F. Alday}
\author[b]{and Gregory P. Korchemsky}

\affiliation[a]{Mathematical Institute, University of Oxford,  Andrew Wiles Building, Radcliffe Observatory Quarter, Woodstock Road, Oxford, OX2 6GG, UK}

\affiliation[b]{Institut de Physique Th\'eorique\footnote{Unit\'e Mixte de Recherche 3681 du CNRS}, Universit\'e Paris Saclay, CNRS, CEA, F-91191 Gif-sur-Yvette}

\abstract{ We revisit the calculation of instanton effects in correlation functions in ${\cal N}=4$ SYM involving the Konishi operator and operators
of twist two. Previous studies revealed that the scaling dimensions and the OPE coefficients of these operators do not receive instanton corrections 
in the semiclassical approximation. We go beyond this approximation and demonstrate that, while operators belonging to the same 
${\cal N}=4$ supermultiplet ought to have the same conformal data, the evaluation of quantum instanton corrections for one operator can be mapped into a semiclassical 
computation for another operator in the same supermultiplet. This observation allows us to compute explicitly the leading instanton correction to the scaling 
dimension of operators in the Konishi supermultiplet as well as to their structure constants in the OPE of  two half-BPS scalar operators. We then use these results, together with crossing symmetry, to determine instanton corrections to scaling dimensions of twist--four operators with large spin.
}

\begin{document}

\maketitle

\section{Introduction}

Recently impressive progress has been achieved in understanding the properties of four-dimensional maximally supersymmetric $\mathcal N=4$ Yang-Mills theory in the planar limit, see \cite{Beisert:2010jr}. 
Thanks to integrability of the theory in this limit, it becomes possible to compute 
 various quantities
for an arbitrary 't Hooft coupling constant. At weak 
coupling, the resulting expressions agree with the results of explicit perturbative calculation whereas at strong coupling they match the 
predictions coming from the AdS/CFT correspondence. Much less is known however about the properties of $\mathcal N=4$ SYM beyond 
the planar limit and in particular about nonperturbative effects induced by instanton corrections. 

The motivation for studying instanton corrections is multifold. Firstly, $\mathcal N=4$ SYM possesses $S-$duality \cite{Montonen:1977sn,Sen:1994yi,Witten:1995zh}, namely,  
invariance under modular $SL(2,\mathbb Z)$ transformations acting on the complexified coupling constant
\begin{equation}\label{tau}
\tau = \frac{\theta}{2\pi} + \frac{4 \pi i}{g^2} \,. 
\end{equation}
Instantons are expected to play a crucial role in restoring the invariance of the spectrum of scaling dimensions under the $S-$duality group. 
Secondly, previous studies revealed a remarkable similarity between instanton corrections to correlation functions in 
$\mathcal N=4$ SYM at weak coupling and the dual supergravity amplitudes induced by D--instantons in type IIB string theory  \cite{Bianchi:1998nk,Dorey:1998xe,Dorey:1999pd}. 
This suggests that the AdS/CFT correspondence can be tested beyond the planar limit to include instanton effects. 
Finally, the crossing symmetry of correlation functions leads to nontrivial constraints for the conformal data of
the theory. They have been used in \cite{Beem:2013qxa}  to derive bounds for the  scaling dimensions of leading twist operators of various spins. 
These bounds are expected to be saturated at fixed points of the $S-$duality group \cite{Beem:2013hha,Alday:2013bha}, were instanton contributions cannot be neglected.

In this paper, we revisit the calculation of  instanton corrections to various correlation functions in $\mathcal N=4$ SYM
at weak coupling.  To compute such corrections we follow the standard approach (see reviews \cite{Dorey:2002ik,Belitsky:2000ws,Bianchi:2007ft}). Namely, 
we decompose all fields into the sum of classical instanton solutions and fluctuations and then integrate out the latter.
In the semiclassical approximation the quantum fluctuations can be neglected and the correlation functions can be
reduced to finite-dimensional integrals over the collective coordinates of instantons
\begin{align}\label{setup}
\vev{O(1) \dots O(n)}_{\rm inst} = \int d\mu_{\rm phys} \e^{-S_{\rm inst}}\, O(1) \dots O(n)\,,
\end{align}
where all fields on the right-hand side are replaced by their expressions on the instanton background.

In the simplest case of the $SU(2)$ gauge group,  the one-instanton solution depends on bosonic collective
coordinates $\rho$ and $x_0$ defining the size of the instanton and its location as well as on $16$ fermionic 
coordinates $\xi_\alpha^A$ and $\bar\eta_{\dot\alpha}^A$ (with $A=1,\dots,4$ and $\alpha,\dot\alpha=1,2$), 
reflecting the invariance of the equations of motion under $\mathcal N=4$ superconformal transformations.
The corresponding integration measure over the collective coordinates for the one-instanton sector in the $SU(2)$ gauge group
is \cite{Bianchi:1998nk}
\begin{align}\label{measure}
 \int d\mu_{\rm phys} \e^{-S_{\rm inst}} = {g^8 \over 2^{34}\pi^{10}}\e^{2\pi i\tau} \int d^4 x_0 \int {d\rho\over\rho^5} \int d^8 \xi
 \int d^8\bar\eta\,.
\end{align}
For the correlation function \re{setup} to be different from zero, the product of operators $O(1) ... O(n)$ should absorb
all $16$ fermion modes. This property can be used to show the vanishing of leading instanton corrections to
various correlation functions.
The relations \re{setup} and \re{measure} can be generalized to the $SU(N)$ gauge group for the one-instanton 
solution \cite{Dorey:1998xe,Dorey:2002ik} and for multi-instanton solutions at large $N$, see \cite{Dorey:1999pd}. 

Applying this approach we can systematically take into account the instanton effects to correlation functions 
and, then, extract the corresponding corrections to the conformal data of the theory. In particular, the instanton corrections 
to the scaling dimensions of operators have the following general 
form \cite{Bianchi:2007ft}: \footnote{It remains unclear, however, whether the scaling dimensions of unprotected operators
receive nonzero corrections running in powers of $\e^{2\pi i(\tau-\bar\tau)}$ due to the contribution of instanton-anti-instanton configurations. 
For some evidence on the string theory side see \cite{Green:2005ba}.
We thank Nick Dorey and Pierre Vanhove for discussions about this point. }
\begin{align}
\gamma_{\rm inst} = \sum_{n\ge 1}  \lr{ \e^{2\pi i n \tau} + \e^{-2\pi i n \bar\tau}} \sum_{k=0}^\infty \gamma_{n,k}(N) g^{2k}  \,,
\end{align}
where the two terms inside the brackets, $\e^{2\pi i n \tau}$ and  $\e^{-2\pi i n \bar\tau}$, describe the leading contribution of $n$ instantons 
and $n$ anti-instantons, respectively, and  perturbative fluctuations produce subleading corrections suppressed by powers of the
coupling constant $g^{2}$.

Previous studies of four-point and two-point correlation functions revealed \cite{Bianchi:1999ge,Arutyunov:2000im,Bianchi:2001cm,Kovacs:2003rt} that for many operators in $\mathcal N=4$ SYM, 
including those with lower bare dimension (Konishi operator) and twist-two operators,
the leading instanton corrections vanish, $\gamma_{n,0}(N)=0$. Going beyond the semiclassical approximation, one can 
envisage two possible scenarios: (i) the instanton corrections vanish to all orders in the coupling constant, $\gamma_{n,k}=0$, due 
to some symmetry, or (ii) the instanton corrections do not vanish, namely $\gamma_{n,k}\neq 0$ for $k\ge K$,  but they are suppressed by a power  of
the coupling constant $g^{2K}$. The first scenario seems to be incompatible with the expected $S-$duality of $\mathcal N=4$ SYM, whereas to test
the second scenario would require to take into account quantum corrections making the calculation of instanton effects extremely
complicated. This explains, in part, why little progress has been made in improving the existing results  over the last decade. 

In this paper we demonstrate, for the first time, that the scaling dimensions of the Konishi operator as well as other members of the corresponding
$\mathcal N=4$ supermultiplet receive instanton corrections at order $O(g^4)$, that is $\gamma_{n,0}=\gamma_{n,1}=0$ but $\gamma_{n,2} \neq 0$. We 
identify the leading nonvanishing correction and compute the corresponding coefficient $\gamma_{n,2}$. We also evaluate the 
three-point correlation function of the Konishi operator and two half-BPS scalar operators and show that
it receives a nonvanishing instanton 
correction at order $O(g^2)$. While operators of the same supermultiplet ought to have the same anomalous dimension (and related OPE coefficients with two half-BPS operators), 
we observe that quantum instanton computations for some operators map to semiclassical instanton computations for others! This allows us to make progress. 

Using these results, we obtain the instanton contribution to the asymptotic behavior of the four-point
correlation function of half-BPS operators in the light-cone limit and, then, 
 employ crossing symmetry to compute the instanton corrections to twist-four operators with high spin. 

The paper is organized as follows. In Sect.~2 we review the conventional instanton calculus in $\mathcal N=4$ SYM and apply it
to compute instanton effects to various correlation functions involving the Konishi operator. 
We show that, due to a different coupling constant dependence of the leading instanton corrections to two- and four-point correlation functions, it is to possible to compute $O(g^4)$
corrections to the scaling dimension of the Konishi operator. Furthermore, we consider anomalous dimensions as well as OPE coefficients (with two half-BPS operators) corresponding to general twist-two operators. 
We show that at order $O(g^2)$ only the OPE coefficient corresponding to the Konishi supermultiplet gets instanton constributions. In Sect.~3 we use this information in order to infer the instanton contribution to the light-cone asymptotics of correlation function of four half-BPS operators. This information, together with crossing symmetry, is then used to compute the instanton corrections to twist-four operators with large spin. 
Concluding remarks are presented in Sect.~4.
Useful definitions are included in two appendices.  

\section{Instanton corrections to correlation functions}

In this Section, we evaluate instanton corrections to two- and three-point correlation functions of various 
operators in $\mathcal N=4$ SYM theory in the semiclassical approximation. As was explained in 
the previous section, the calculation amounts to evaluating the product of operators in the background of
instantons and integrating the resulting expression over the collective coordinates.

\subsection{Instanton in $\mathcal N=4$ SYM}

The Lagrangian of $\mathcal N=4$ super Yang-Mills theory describes a gauge field $A_\mu$,  (anti)gaugino 
fieds  $\lambda_\alpha^A$ and $\bar\lambda^{\dot\alpha}_A$, as well as scalars  $\phi^{AB}$ satisfying the
reality condition $\bar\phi_{AB} = \frac12\epsilon_{ABCD} \phi^{CD}$ 
\begin{align}\notag\label{La}
L {}& = {1\over g^2} \tr \Big\{ - \frac1{16} F_{\alpha\beta}^2 - \frac1{16} F_{\dot\alpha\dot\beta}^2 - \frac14 D^\alpha_{\dot\alpha} \phi^{AB} D_\alpha^{\dot\alpha} \bar\phi_{AB} 
- 2 i \bar\lambda_{\dot\alpha A} D^{\dot\alpha\beta} \lambda_\beta^A+ \sqrt{2} \lambda^{\alpha A} [\bar\phi_{AB},\lambda_\alpha^B]
\\  
{}& 
- \sqrt{2} \bar\lambda_{\dot\alpha A} [\phi^{AB},\bar\lambda_B^{\dot\alpha}]+\frac18 [\phi^{AB},\phi^{CD}][\bar\phi_{AB},\bar\phi_{CD}] \Big\}
 +  i{\theta\over 8\pi^2}\tr \Big\{ \frac1{16} F_{\alpha\beta}^2- \frac1{16} F_{\dot\alpha\dot\beta}^2\Big\}\,.
\end{align}
Here we used spinor notations (see Appendix~A for our conventionts) and denoted by
$F_{\alpha\beta}$ and $F_{\dot\alpha\dot\beta}$ the (anti) self-dual part of gauge field strength tensor $F_{\mu\nu} = -i [D_\mu,D_\nu]$ 
with covariant derivative $D_\mu = \partial_\mu + i [A_\mu, ]$ and $D_{\alpha\dot\alpha } =D_\mu (\sigma^\mu)_{\alpha\dot\alpha }$.  
All fields take value in the $SU(N)$ algebra, e.g. $A_\mu= A_{\mu}^a\, T^a$, with the generators satisfying 
$[T^a, T^b] = i f^{abc} T^c$ and normalized as $\tr(T^a T^b) = \frac12 \delta^{ab}$.

By definition, the instanton is a solution to the classical equations of motion. Due to our choice of normalisations in the Lagrangian \re{La}, 
all elementary fields in the instanton background, $A_\mu$, $\lambda_\alpha^A$, 
$\bar\lambda^{\dot\alpha}_A$ and $\phi^{AB}$, are independent of the coupling constant. Their explicit expressions for the $SU(N)$ gauge group 
are rather complicated and only few terms in their expansion in powers of $8N$ fermionic collective modes are currently known 
\cite{Dorey:2002ik,Belitsky:2000ws,Bianchi:2007ft}. Significant
simplification occurs however for the $SU(2)$ gauge group. In this case, the general one-instanton
solution can be obtained by applying $\mathcal N=4$ superconformal transformations to a special solution to the equations motion
\begin{align}\label{bare}
A_\mu^{(0)} =  2{\eta_{\mu\nu}^a (x-x_0)^\nu T^a \over (x-x_0)^2+\rho^2} \,,\qquad \phi^{AB,(0)}=\lambda_\alpha^{A,(0)}=\bar\lambda^{\dot\alpha,(0)}_A=0\,,
\end{align}
where $A_\mu^{(0)}$ is the well-known one-instanton solution in pure Yang-Mills theory. It depends on the collective coordinates $\rho$ and 
$x_0$ defining the size and the position of the instanton, respectively. Here $\eta_{\mu\nu}^a$ are the 't Hooft symbols and the $SU(2)$ generators are related to Pauli matrices  $T^a=\sigma^a/2$. 

The field configuration \re{bare} is annihilated by half of the superconformal generators 
$\bar Q^{\dot\alpha A}$ and $S_\alpha^A$. Applying the remaining superconformal transformations  $\exp\lr{\xi_\alpha^A Q^\alpha_A+\bar\eta^{\dot\alpha A}\bar S_{\dot\alpha A}}$ to \re{bare}, we obtain a solution to the full classical equations of motion that depends on $16$ fermionic collective coordinates, $\xi_\alpha^A$ and $\bar\eta^{\dot\alpha A}$. By virtue of the $SU(4)$ $R-$symmetry, the resulting expression for the instanton
configuration takes the following form
\begin{align}\notag\label{dec}
{}& A_\mu =A_\mu^{(0)}+A_\mu^{(4)}+\dots + A_\mu^{(16)}\,,
\\[2mm]\notag
{}& \lambda^{\alpha A} = \lambda^{\alpha A, (1)} +  \lambda^{\alpha A, (5)}+\dots + \lambda^{\alpha A, (13)}\,,
\\[2mm]\notag
{}& \phi^{AB} = \phi^{AB,(2)} + \phi^{AB,(6)} + \dots + \phi^{AB,(14)} \,,
\\[1.2mm]
{}& \bar\lambda_{\dot\alpha A} =\bar\lambda_{\dot\alpha A}^{(3)} + \bar\lambda_{\dot\alpha A}^{(7)} + \dots + \bar\lambda_{\dot\alpha A}^{(15)} \,,
\end{align}
where $A_\mu^{(n)}$ denotes the component of the gauge field that is homogenous in $\xi_\alpha^A$ and $ \bar \eta^{\dot\alpha A}$ of degree
$n$ and similar for other fields. 

The leading terms of the expansions \re{dec} have been worked out in Ref.~\cite{Belitsky:2000ws}. For our purposes we will also need subleading terms. 
Their direct calculation is more involved, e.g. finding $\phi^{AB,(6)}$ amounts to applying  $Q-$ and $\bar S-$transformations to \re{bare} 
subsequently six times. There is, however, a shortcut that simplifies the calculation significantly. Namely, the subleading corrections depend
on the instanton field $A_{\alpha\dot\alpha}^{(0)}= i A_\mu^{(0)}(\sigma^\mu)_{\alpha\dot\alpha}$ and fermion collective coordinates, 
$\xi_\alpha^A$ and $\bar\eta^{\dot\alpha A}$. It turns out that the requirement for the fields \re{dec} to have the correct properties with respect to 
conformal symmetry, $R-$symmetry and gauge transformations, fixes their general form up to a few constants. The latter can be determined by requiring the 
fields \re{dec} to satisfy the classical equations of motion derived from \re{La}.
Going through the calculation we have found the expressions for the subleading corrections to gauge field $A_{\alpha\dot\alpha}^{(4)}= i A_\mu^{(4)}(\sigma^\mu)_{\alpha\dot\alpha}$ and scalar $\phi^{(6),AB}$. Together with the leading correction  $\phi^{(2),AB}$
they are given by 
\begin{align}\notag\label{phi2}
{}& A^{(4)}_{\alpha\dot\alpha}  
= - \frac1{12}\epsilon_{ABCD} \zeta_\alpha^{A} \zeta^{\beta B} (\zeta  D_{\beta\dot\alpha} F\zeta)^{CD}
- \frac12\epsilon_{ABCD}  \zeta_\alpha^{A} \bar\eta_{\dot\alpha}^B (\zeta F\zeta)^{CD}\,,
\\[2mm]\notag
{}& \phi^{(2),AB} =  
{1\over \sqrt{2}} (\zeta F \zeta)^{AB} \,, 
\\  
{}& \phi^{(6),AB} = - {1\over 20\sqrt{2}}\epsilon_{CDEF}   (\zeta^2)^{AC} (\zeta^2)^{BD} (\zeta F^2\zeta)^{EF}\,,
\end{align}
where we have introduced a short-hand notation for a particular linear $x-$dependent combination of fermionic modes
\begin{align}\label{xi}
\zeta_\alpha^A(x)=\xi_\alpha^A + x_{\alpha\dot\alpha}\bar\eta^{\dot\alpha A} \,,
\end{align}
and for various Lorentz contractions of fermion modes 
\begin{align}\notag\label{xi2-def}
{}& (\zeta^2)^{AB} = (\zeta^2)^{BA} =  \zeta^{\beta A}\epsilon_{\beta\gamma} \zeta^{\gamma B}\,,
\\ \notag
{}&(\zeta F \zeta)^{AB} =-(\zeta F \zeta)^{BA} =   \zeta^{\alpha A} F_{\alpha\beta} \zeta^{\beta B}\,,
\\  
{}&(\zeta F^2 \zeta)^{AB} =-(\zeta F^2 \zeta)^{BA} =   \zeta^{\alpha A} F_{\alpha\beta} \epsilon^{\beta\delta}F_{\delta\gamma} \zeta^{\gamma B}\,.
\end{align}
Here $F_{\alpha\beta}=F_{\beta\alpha}$ is the strength tensor for the $SU(2)$ instanton
\footnote{We recall that the anti self-dual part of the instanton strength tensor vanishes, $F_{\dot\alpha\dot\beta}=0$.}
\begin{align}\label{F-inst}
(F_{\alpha\beta})_i{}^j  ={8\rho^2\over [(x-x_0)^2+\rho^2]^2} \big( \epsilon_{i\alpha}\delta^j_{\beta}+ \epsilon_{i\beta}\delta^j_{\alpha} \big)\,,
\end{align}
where $i$ and $j$ are the $SU(2)$ indices. 

A peculiar feature of $\phi^{(6),AB}$ in \re{phi2} is that it depends on the fermionic modes only through the variable $\zeta$ defined in \re{xi}. This is not
the case however for the gauge field $A^{(4)}_{\alpha\dot\alpha}$. The difference is due to different transformation properties of fields
under conformal transformations. By virtue of superconformal invariance, the action of $\mathcal N=4$ SYM 
evaluated on the instanton configuration \re{dec} does not depend on the fermionic modes $\xi$ and $\bar\eta$ and is given by
\begin{align}
S_{\rm inst}=
\int d^4 x \, L(x) = {8\pi^2\over g^2} -i\theta  = - 2\pi i \tau\,,
\end{align}  
where $\tau$ is the complex coupling constant \re{tau}.

\subsection{Normalization of operators}

Later in this section we shall compute the leading instanton corrections to scaling dimensions and OPE coefficients of various
composite gauge invariant operators built from scalar fields $\phi^{AB}$ and $\bar\phi_{AB}=\frac12\epsilon_{ABCD}\phi^{CD}$.

Before doing this, we have to carefully examine the normalization of the operators. The reason for this is that, due to our definition 
of the Lagrangian \re{La}, the free scalar propagator depends on the coupling constant
\begin{align}\notag
{}& \vev{\bar\phi_{AB}^a (1) \phi^{b,CD}(2)} = g^2  \delta^{ab} \lr{\delta_A^C\delta_B^D- \delta_A^D\delta_B^C}D(x_{12}) \,,
\\[2mm]
{}& \vev{ \phi^{a,AB}(1) \phi^{b,CD}(2)} = g^2 \delta^{ab}\epsilon^{ABCD}D(x_{12})\,,
\end{align}
where we used a shorthand notation for $x_{12}=x_1-x_2$ and $D(x) =1/(4\pi^2 x^2)$. Taking this into account, we
 define the simplest scalar operators of bare dimension $2$
\begin{align}\notag\label{K-def}
{}&
 O_{\bf 20'}(x,Y) ={1\over g^2} Y_{AB} Y_{CD} \tr(\phi^{AB}\phi^{CD})\,,
\\ 
{}& K(x) = {1\over g^2} \tr (\bar\phi_{AB} \phi^{AB}) \,,
\end{align}
where $Y_{AB}$ is an antisymmetric $SU(4)$
tensor satisfying $\epsilon^{ABCD} Y_{AB} Y_{CD}=0$. It was introduced to project the product of two scalar fields onto the irreducible
$SU(4)$ representation $\bf 20'$. The half-BPS operator $\mathcal O_{\bf 20'}$ is annihilated by half of the $\mathcal N=4$ supersymmetries and its 
scaling dimension is protected from quantum corrections. The Konishi operator $K$ is the simplest unprotected operator.

It is also convenient to introduce the following operator of bare dimension $4$
\begin{align}\label{K'-def}
  K'(x) ={1\over g^4}\tr ( [Z,X][Z,X] )\,,
\end{align}
where we used the standard notation for complex scalar fields $Z=\phi^{14}$ and $X=\phi^{24}$. 
The operator $K'$ is a supersymmetric descendant of the Konishi operator, $K' \sim \delta_{\bar Q}^2 \delta_Q^2 K$, and, as 
a consequence, its anomalous dimension and OPE coefficient with two half-BPS operators coincide with those of the Konishi operator.
 
The definition of the operators in \re{K-def} and \re{K'-def} involves inverse powers of the coupling constant, one per each scalar field. 
They were introduced in order to ensure that the correlation functions scale as $O(g^0)$ in the Born approximation. Indeed
\begin{align}\notag\label{Born}
 {}&  \vev{K(1)   K(2)}_{\rm Born} = 12 (N^2-1) D^2(x_{12})\,,
\\[2mm]
\notag
 {}&  \vev{K'(1) \bar{K'}(2)} _{\rm Born}= \frac34 N^2(N^2-1) D^4(x_{12})\,,
\\
 {}&  \vev{O_{\bf 20'}(1) O_{\bf 20'}(2)}_{\rm Born} =\frac12 (y_{12}^2)^2 (N^2-1) D^2(x_{12})\,,
\end{align} 
where $y_{12}^2\equiv \epsilon^{ABCD} Y_{1,AB} Y_{2,CD}$.

The presence of different powers of the coupling constant in the definition of the operators \re{K-def} and \re{K'-def} leads to important consequences
for instanton corrections to the correlation functions. Since the fields in the instanton background  \re{dec} do not depend on the
coupling constant, the dependence on $g^2$ of the correlation function in the semiclassical approximation comes solely from 
the $SU(2)$ integration measure over the moduli of instantons \re{measure} and the from powers of $1/g^2$ accompanying each operator.
In this way, we find 
\begin{align}\label{KK}\notag
{}& \vev{K(1) K(2)}_{\rm inst}  = O(g^4\e^{2\pi i\tau})\,, 
\\[2mm]
{}& \vev{K'(1) \bar K'(2)}_{\rm inst}  = O(\e^{2\pi i\tau})\,,
\end{align}
where the extra factor of $g^4$ in the first relation arises due to a different power of $1/g^2$ in the definitions \re{K-def} and \re{K'-def}.

In general, two-point correlation functions develop logarithmic singularities and generate corrections to anomalous dimensions of operators. 
Then, assuming that the correlation functions \re{KK} are different from zero, we would deduce that instanton corrections to anomalous 
dimensions of operators $K$ and $K'$ should have different dependence on the coupling constant: $\gamma_{K'} = O(\e^{2\pi i \tau})$ whereas 
$\gamma_{K} = O(g^4 \e^{2\pi i \tau})$. However, this contradicts the fact that the two operators belong to the same supermultiplet and,
therefore, their anomalous dimensions have to coincide. In other words, the superconformal symmetry dictates that $\gamma_{K'}$
has to vanish in the semiclassical approximation.~\footnote{Indeed, as we show in the next subsection, the correlation functions involving
the operator $K'$ do vanish in the semiclassical approximation.}
To get a nonzero result for $\gamma_{K'}$ we have to go beyond this approximation and
take into account quantum fluctuation around instanton onfigurations. The calculation of quantum corrections to $\vev{K'(1) \bar K'(2)}_{\rm inst}$ is way
more complicated but the resulting expression for $\gamma_{K'}$ ought to match $\gamma_{K}= O(g^4 \e^{2\pi i \tau})$ obtained in the 
semiclassical approximation. We shall use this observation to compute the leading instanton correction to the scaling dimension of
the Konishi operator in Sect.~\ref{sect:K}.

To observe another interesting feature of the Konishi operator, we examine the coupling dependence of the following correlation functions
\begin{align} \notag\label{3pt}
{}& \vev{O_{\bf 20'}(1) O_{\bf 20'}(2) K(3)}_{\rm inst}  = O(g^2 \e^{2\pi i\tau})  \,,
\\[2mm]
{}& \vev{O_{\bf 20'}(1) O_{\bf 20'}(2)O_{\bf 20'}(3) O_{\bf 20'}(4)}_{\rm inst} = O(\e^{2\pi i\tau})  \,,
\end{align}
where each operator brings in the factor of $1/g^2$ multiplied by the factor of  $g^8 \exp(2\pi i \tau)$ coming from the $SU(2)$ integration 
measure. Performing conformal partial wave expansion of the four-point correlation function in the second 
relation in \re{3pt} we can identify the contribution of the operators whose anomalous dimensions and/or OPE coefficients scale as $O(\e^{2\pi i\tau})$. At the same time,
as follows from the first relation in \re{3pt}, the OPE coefficient of the Konishi operator scales as $O(g^2 \e^{2\pi i\tau})$ and, therefore,
it provides a vanishing contribution to the four-point correlation function of half-BPS operators to order $O(\e^{2\pi i\tau})$, in 
agreement with findings of Refs.~\cite{Bianchi:1999ge,Arutyunov:2000im}. However, we can also turn the logic around and use the first relation in \re{3pt}
to predict the leading $O(g^2 \e^{2\pi i\tau})$ contribution of the Konishi supermultiplet to the four-point correlation function! A direct calculation of 
such correction would require taking into account quantum corrections. 
 
\subsection{Instanton profile of operators}

As the first step, we evaluate the operators \re{K-def} and  \re{K'-def}  in 
the instanton background for the $SU(2)$ gauge group. We start with the Konishi operator \re{K-def} and replace scalar fields by their expressions \re{dec}. This leads to
\begin{align}\label{K-1}
K ={1\over g^2} \tr (\phi^{AB} \bar\phi_{AB}) ={1\over g^2}\left[ K^{(4)}+  K^{(8)}  + K^{(12)}   + K^{(16)} \right]\,,
\end{align} 
where $K^{(n)}$ denotes the contribution containing $n$ fermion modes. Notice that $K^{(n)}$ is independent
of the coupling constant. Since the scalar field has at least
two fermion modes, the expansion starts with $K^{(4)}$. By virtue of the $SU(4)$ symmetry, the
number of fermion modes in the subsequent terms of the expansion increases by four units.
The last term of the expansion contains the maximal number of fermion modes. 
We find, in agreement with \cite{Bianchi:2001cm}, that the first term on the right-hand side of \re{K-1} vanishes
\begin{align} 
{}& K^{(4)}(x) = \tr \big[\phi^{(2),AB} \bar\phi^{(2)}_{AB}\big]  \sim \epsilon_{ABCD} (\zeta^2)^{AB} (\zeta^2)^{CD} = 0 \,,
\end{align}
where $\zeta=\xi+ x\bar\eta$ is a linear combination of fermion modes defined in \re{xi}.
Here in the second relation we took into account that $(\zeta^2)^{AB} = \zeta^{\alpha A} \zeta_{\alpha}^B$ 
is symmetric with respect to $SU(4)$ indices. As a result, the expression for the Konishi operator contains
eight fermion modes at least and the leading term is given by
\begin{align}\label{K-3}
  K^{(8)}(x)= 2  \tr \big[\phi^{(2),AB} \bar\phi^{(6)}_{AB}\big]\ = -3^2 \times 2^{15} \times {\rho^6\over [\rho^2+(x-x_0)^2]^6} 
  [\zeta(x)]^8 \,,
\end{align}
where $\zeta^8 =\prod_{A,\alpha} \zeta^{\alpha A}$ is the product of eight fermion modes. Expressions for higher
components of \re{K-1} are more complicated but we do not need them for our purposes.
 
Let us consider the half-BPS operator $O_{\bf 20'}(x,Y)$ defined in \re{K-def}. Since this operator is annihilated by half of the
$\mathcal N=4$ supercharges, it depends on four $\xi$ and four $\bar\eta$ fermion modes. As a consequence, its expansion 
in powers of Grassmann variables is shorter as compared with \re{K-1}
\begin{align}\label{O20-dec}
O_{\bf 20'}(x,Y) = {1\over g^2}Y_{AB} Y_{CD} \tr(\phi^{AB}\phi^{CD})
 ={1\over g^2}\left[ O^{(4)}_{\bf 20'}+O^{(8)}_{\bf 20'}\right],
\end{align}
where the two terms on the right-hand side involve four and eight fermion modes respectively, and are given by
\begin{align}\notag\label{O-com}
{}& O^{(4)}_{\bf 20'} = Y_{AB} Y_{CD} \tr(\phi^{(2),AB}\phi^{(2),CD})\,,
\\[2mm]
{}& O^{(8)}_{\bf 20'} = 2 Y_{AB} Y_{CD}\tr(\phi^{(2),AB}\phi^{(6),CD})\,.
\end{align}
As was already mentioned, the scalar fields $\phi^{(2),AB}$ and $\phi^{(6),AB}$ depend on fermion modes
only through their linear combination \re{xi} and the same is obviously true for the components \re{O-com}.
This property alone implies that $O^{(8)}_{\bf 20'}$ has to vanish. Indeed, if $O^{(8)}_{\bf 20'}$ were
different from zero, it would be proportional to $\zeta^8=(\xi + x \bar\eta)^8$ and, therefore, would
contain terms with more than four $\xi$'s and four $\bar\eta$'s,
in contradiction with half-BPS condition. The relation $O^{(8)}_{\bf 20'}=0$ can be also verified by a direct 
calculation.\footnote{Indeed, $\tr(\phi^{(2),AB}\phi^{(6),CD})$ is proportional to 
$\zeta^8 \epsilon^{ABCD}$ and, therefore, $O^{(8)}_{\bf 20'}\sim Y_{AB} Y_{CD} \epsilon^{ABCD}=0$.}
Thus, the operator $O_{\bf 20'}(x,Y)$ contains exactly $4$ fermion $\zeta-$modes and is given by 
\begin{align}\label{O-inst}
O_{\bf 20'}(x,Y)  {}& =  {1\over g^2}{128 \rho^4\over [\rho^2+(x-x_0)^2]^4}\times Y_{AB} Y_{CD} (\zeta^2)^{AC} (\zeta^2)^{BD} \,.
\end{align}
This feature allows us to verify the known properties of correlation functions of  half-BPS operators.

We recall that in order for a correlation function to be different from zero, the product
of operators should involve terms containing sixteen fermion modes. Since the product of two and three
half-BPS operators have $8$ and $12$ fermion modes, respectively, the corresponding correlation functions
$\vev{O_{\bf 20'}(1) O_{\bf 20'}(2) }$ and $\vev{O_{\bf 20'}(1) O_{\bf 20'}(2)O_{\bf 20'}(3) }$ do not receive
instanton corrections  in the semiclassical approximation. This result is in agreement with the known
fact that the above mentioned correlation functions are protected from quantum corrections and are given by their Born level
expressions. The simplest correlation function that receives instanton correction involves four half-BPS
operators. We shall return to this correlation function in Sect.~\ref{bootstrapping}.

Let us finally examine the operator $K'$ defined in \re{K-def}. Replacing the scalar fields by their expressions 
\re{dec}, we find in a similar manner
\begin{align}\label{K'}
 K' ={1\over g^4}\tr \left( [Z,X][Z,X] \right) 
 ={1\over g^4}\left[K^{'(8)}  + K^{'(12)}   + K^{'(16)} \right]\,,
\end{align} 
where the lowest term involves eight fermion modes, 
\begin{align} \notag\label{K'12}
{}& K^{'(8)} = \tr \left( [Z^{(2)},X^{(2)}][Z^{(2)},X^{(2)}] \right),
\\[2mm]
{}& K^{'(12)} = 2\tr \left( [Z^{(6)},X^{(2)}][Z^{(2)},X^{(2)}] \right) + 2\tr \left( [Z^{(2)},X^{(6)}][Z^{(2)},X^{(2)}] \right).
\end{align}
As in the previous case, the dependence on fermion modes enters into these expressions through their linear
combination $\zeta_\alpha^{A}$ defined in \re{xi}. According to \re{phi2}, the scalar field $\phi^{(2),AB}$ is given by
the product of two fermion modes $\phi^{(2),AB}\sim \zeta^A\zeta^B$. Taking into account that $Z=\phi^{14}$ and 
$X=\phi^{24}$, we obtain that $K^{'(8)} \sim (\zeta^{4 \alpha} \zeta_\alpha^{4})^2 = 0$, which is consistent with the findings of \cite{Kovacs:2003rt}.
In a similar manner, we can show that
$K^{'(12)}\sim (\zeta^{4 \alpha} \zeta_\alpha^{4}) \zeta_\beta^{4} =0$. The top component $K^{'(16)}$ depends on all $16$
fermion $\xi-$ and $\bar\eta-$modes. However their product $ \xi^8\,\bar\eta^8$ is the $SU(4)$ singlet 
and can not contribute to $K^{'(16)}$ due to mismatch of the $SU(4)$ quantum numbers leading to $ K^{'(16)}=0$.
We therefore conclude that $K'=0$ in the instanton background and, as a consequence, all correlation
functions involving this operator vanish in the semi-classical approximation.
 
\subsection{Instanton corrections to Konishi operator} \label{sect:K}
 
 We are now ready to evaluate the leading instanton corrections to correlation functions involving the Konishi operator
 for the $SU(2)$ gauge group.
 We start with the two-point function 
\begin{align}\notag
{}&\vev{K(1) K(2)}_{\rm inst} 
 =  \int d\mu_{\rm phys}\, \e^{-S_{\rm isnt}}K(1)K(2)
={\e^{2\pi i \tau}\over g^4} \int d\mu_{\rm phys}\,K^{(8)}(1)K^{(8)}(2)\,,
\end{align}
where the $SU(2)$ integration measure $\int d\mu_{\rm phys}$ is defined in \re{measure}. Here
in the first relation we replaced operators by their instanton profile \re{K-1} and in the second relation retained
terms involving $16$ fermion modes. The terms with higher number of fermion modes do not contribute.
Replacing $K^{(8)}$ with \re{K-3} we get
\begin{align}\notag
\vev{K(1)K(2)}_{\rm inst}  {}& = 
  {81\over 16 \pi^{10}} g^{4}\e^{2\pi i\tau}\int d^4 x_0 \int {d\rho\over \rho^5} 
   {\rho^{12}\over (\rho^2+x_{10}^2)^6(\rho^2+x_{20}^2)^6}
\\
{}&  \times 
  \int d^8 \xi\int d^8 \bar\eta \,\prod_{A=1}^4 \zeta_1^{A}(x_1)   \zeta_2^{A}(x_1) \zeta_1^{A}(x_2)   \zeta_2^{A}(x_2)\,,
\end{align}
where $\zeta_\alpha^A(x)=\xi_\alpha^A + x_{\alpha\dot\alpha}\bar\eta^{\dot\alpha A}$. Integration over
Grassmann variables in the second line yields $(x_{12}^2)^4$ leading to
\begin{align}\label{KK-int}
\vev{K(1) K(2)}_{\rm inst}  {}& = 
 {81\over 16 \pi^{10}}  g^{4}  \e^{2\pi i\tau}\int d^4 x_0 \int {d\rho\over \rho^5} 
 {(x_{12}^2)^4 \rho^{12} \over (\rho^2+x_{10}^2)^6(\rho^2+x_{20}^2)^6}\,,
\end{align}
where the integration is over the size and position of the instanton.
We verify that this expression has the expected dependence \re{KK} on the coupling constant.

The integral on the right-hand side of \re{KK-int} develops a logarithmic divergence that comes from integration over 
instantons of small size, $\rho\to 0$, located close to one of the operators, $|x_{10}|\to 0$ or $|x_{20}|\to 0$. It indicates
that the instanton corrections modify the scaling dimension of the operator. It is convenient to regularize the integral by
modifying the integration measure over $x_0$
\begin{align}
\int d^4 x_0 \quad \to\quad 
\int d^{4-2\epsilon} x_0\,.
\end{align}
The resulting integral in \re{KK-int} is well-defined for $\epsilon<0$ and develops a simple pole 
as $\epsilon\to 0$.  
Combining \re{KK-int} with the Born term \re{Born} (evaluated for the $SU(2)$ gauge group) we obtain 
\begin{align} 
\vev{K(1) K(2)} ={36 \over (4\pi^2)^2 (x_{12}^2)^{2}}\bigg[1 - {9\over 5 \epsilon} (x_{12}^2)^{-\epsilon}\lr{g^{2}\over 4\pi^2}^2  \e^{2\pi i\tau}  \bigg]\,.
\end{align}
Following a standard procedure, we apply the dilatation operator $\sum_i (x_i\partial_i)$ to both sides of this relation, and find that
the correction to the scaling dimension of $K$ is given by the residue at the pole. Thus, we conclude that the leading instanton correction
to the anomalous dimension of the Konishi operator in the $SU(2)$ gauge group is given by
\begin{align}\label{anom-K}
\gamma_{K} = - {9\over 5}\lr{g^{2}\over 4\pi^2}^2 \lr{ \e^{2\pi i\tau} + \e^{-2\pi i\bar \tau} } \,,
\end{align}
where we have added a complex conjugated term to take into account the contribution from the anti-instanton. Notice that $\gamma_{K}$ 
has negative sign. The result \re{anom-K} holds for the $SU(2)$ gauge group. Its generalization to $SU(N)$ gauge 
group will be discussed in Sect.~\ref{sec:SU(N)}. 
 
For the three-point function of a Konishi operator and two half-BPS operators we can proceed analogously. We find
\begin{align}\notag\label{OOK-int}
 \vev{O_{\bf 20'}(1) O_{\bf 20'}(2) K(3)}_{\rm inst} 
{}&  ={\e^{2\pi i\tau}\over g^2} \int d\mu_{\rm phys} \, O_{\rm 20'}(1) O_{\rm 20'}(2) K^{(8)}(3)
\\
{}& =  - {9\over 32 \pi^{10}}g^{2}  \e^{2\pi i \tau}\times I_{\rm B} \times I_{\rm F}\,,
\end{align}
where in the second relation we replaced operators by their expressions on the instanton background, Eqs.~\re{K-3} and \re{O-inst}, and introduced
a short-hand notation for the integrals over bosonic and fermion collective coordinates  
\begin{align}  \notag
I_{\rm B}  {}&= \int d^4 x_0 \int {d\rho\over \rho^5}{\rho^{14}\over (\rho^2+x_{10}^2)^4(\rho^2+x_{20}^2)^4(\rho^2+x_{30}^2)^6}\,,
\\[2mm]\notag
I_{\rm F}  {}&= \int d^8 \xi\int d^8 \bar\eta\,  Y_{1,AB} Y_{1,CD} (\zeta^2(x_1))^{AC} (\zeta^2(x_1))^{BD}
\\[-2mm]
{}&\qqqquad \times Y_{2,EF} Y_{2,KL} (\zeta^2(x_2))^{EK} (\zeta^2(x_2))^{FL}\prod_{A=1}^4 \zeta_1^{A}(x_3)   \zeta_2^{A}(x_3)\,,
\end{align}
with $\zeta^2(x)$ given by \re{xi2-def} and \re{xi}.
Both integrals are well-defined and their dependence on $x-$ and $Y-$variables is uniquely fixed by conformal and $R-$symmetry,
respectively. Going through the calculation we get
\begin{align}\label{IB}
I_{\rm B} = {\pi^2\over 90}{1\over (x_{13}^2x_{23}^2)^3 x_{12}^2}
\,,\qqqquad
I_{\rm F}  = 9  (x_{13}^2 x_{23}^2)^2  (y_{12}^2)^2\,,
\end{align} 
where $y_{12}^2$ is defined in \re{Born}.  Plugging \re{IB} into \re{OOK-int} we obtain the final result for the instanton contribution. Taking into account the anti-instanton contribution and combining this with the Born term we obtain 
\begin{align}\label{OOK-res}
 \vev{O_{\bf 20'}(1) O_{\bf 20'}(2) K(3)} 
=  {6\,(y_{12}^2)^2\over (4\pi^2)^3 x_{12}^2 x_{23}^2 x_{31}^2}\bigg[1-
 {3g^2\over 10\pi^2}   \lr{\e^{2\pi i \tau}+\e^{-2\pi i\bar  \tau}} \bigg]\,.
\end{align}
We observe that, in agreement with \re{anom-K}, the three-point function does not receive logarithmically divergent corrections due 
to anomalous dimension of the Konishi operator. The latter corrections scale as $O(g^4 \e^{2\pi i\tau})$ and are subleading 
in \re{OOK-res}. As a consequence, the leading instanton correction to the OPE coefficient 
$C_K=\vev{O_{\bf 20'} O_{\bf 20'} K}/(\vev{O_{\bf 20'} O_{\bf 20'}}\vev{K K}^{1/2})$ describing the contribution
of the Konishi (super)multiplet to the product of operators $O_{\bf 20'}(1) O_{\bf 20'}(2)$ differs from the Born level result 
$C^{(0)}_{K}$ by the same factor
that enters the right-hand side of \re{OOK-res} 
\footnote{Here we took into account that the two-point correlation
function of $K$ does not receive instanton correction at order $O(g^2{\e^{2\pi i \tau}})$, see \re{KK}. }
\begin{align}\label{c-K}
C_{K}/C^{(0)}_{K} =  1-
 {3 g^2\over 10\pi^2} \lr{\e^{2\pi i \tau}+\e^{-2\pi i\bar  \tau}}  \,.
\end{align}

\subsection{Instanton corrections to twist-two operators} 

The twist-two operators provide the leading contribution to four-point correlation functions in the light-cone limit $x_{12}^2\to 0$. In
$\mathcal N=4$ SYM these operators belong to the same supermultiplet which allows us to restrict our consideration to a particular
twist-2 operator  
\begin{align}\label{tw2}
\mathcal O_S= {1\over g^2} \tr ( Z D_+^{S} Z ) + \dots\,,
\end{align}
which is built from the complex scalar field $Z=\phi^{14}$ and the light-cone component of the covariant derivative $D_+=(nD)$ (with $n_\mu^2=0$).
The ellipses denote similar terms with covariant derivatives
distributed between the two scalar fields, their form fixed by conformal symmetry. For $S=0$ the operator \re{tw2} coincides with the half-BPS
operator $O_{\bf 20'}(x,Y)$ for a special choice of $Y-$variables. For $S=2$ it is a superconformal
descendant of the Konishi operator $K$. As in the case of Konishi operator, the additional factor of $1/g^2$ was introduced in \re{tw2} 
in order for the two-point correlation function of $\mathcal O_S$ to scale as $O(g^0)$ in the Born approximation.

In order to study instanton corrections to $\mathcal O_S$ it is convenient to switch to nonlocal light-ray operators  
\begin{align} \label{O-LC}
\mathbb O(z) {}&= {1\over g^2} \tr\left[Z(0) E(0,z) Z(nz) E(z,0)\right]
  =\sum_{S\ge 0} {z^S\over S!}\big[ O_S(0) + \dots \big]\,,
\end{align}
where two scalar fields are separated along the light-ray defined by a null vector $n$ and light-like Wilson lines were introduced to 
restore gauge invariance,
\begin{align}
E(z_1,z_2) = P\exp\lr{i \int_{z_1}^{z_2} dt\,  n \cdot A(nt)}\,.
\end{align}
The light-ray operator $\mathbb O(z)$ serves as a generating function of twist-two operators, which are generated by expanding in powers 
of the light-cone separation $z$. The dots on the right-hand side of \re{O-LC} denote the contribution 
of the conformal descendants of the form  $z^\ell  (n\partial)^\ell O_S(0)$.

To find instanton corrections to the three-point function $\vev{O_{\bf 20'}(1) O_{\bf 20'}(2) O_S(0)}$, we can first evaluate 
$\vev{O_{\bf 20'}(1) O_{\bf 20'}(2) \mathbb O(z)}$ and, then, apply \re{O-LC} in order to project it onto the
contribution of the operator $O_S(0)$. As before, we have to examine the product of operators in the background of the
instanton and identify the contributions involving $16$ fermion modes. Since
the product of two half-BPS operators contains $8$ modes, the remaining $8$ modes should be soaked up by 
$\mathbb O(z)$. Replacing $Z=Z^{(2)} + Z^{(6)} + \dots$ and $A=A^{(0)} + A^{(4)} +\dots$ in \re{O-LC} we find the
corresponding contribution is given by
\begin{align}\notag\label{3terms}
\mathbb O^{(8)}(z) {}& =\tr\Big[Z^{(2)}(0) E^{(0)}(0,z)  Z^{(6)}(nz)E^{(0)}(z,0)  
+ Z^{(6)}(0) E^{(0)}(0,z) Z^{(2)}(nz)E^{(0)}(z,0) 
\\[1.6mm]\notag
 {}& +i\int_0^z dt \, Z^{(2)}(0) E^{(0)}(0,t) \,n \cdot A^{(4)}(nt) E^{(0)}(t,z) Z^{(2)}(nz)E^{(0)}(z,0) 
 \\ 
 {}& +i\int^0_z dt  \,Z^{(2)}(0)E^{(0)}(0,z) Z^{(2)}(nz)E^{(0)}(z,t) \, n \cdot A^{(4)}(nt) E^{(0)}(t,0) \Big]\,,
\end{align}
where the subscript indicates the number of fermion zero modes and  $E^{(0)}(z_1,z_2)$
depends on $A^{(0)}$. 

Taking into account \re{phi2} and recalling that $Z=\phi^{14}$, we can evaluate $\mathbb O^{(8)}(z)$ explicitly. 
Going through a lengthy calculation (the details are presented in \cite{Alday:2016jeo}) we obtain
\begin{align}\notag\label{OO-nonloc}
{}& \vev{O_{\bf 20'}(1) O_{\bf 20'}(2) \mathbb O(z)}_{\rm inst}  = {\e^{2\pi i\tau}\over g^2} \int d\mu_{\rm phys} \, O_{\rm 20'}(1) O_{\rm 20'}(2)\mathbb O^{(8)}(z)
\\
{}&\qquad = -{27\over 32\pi^{10}} 
 g^{2}  \e^{2\pi i \tau} y_{12}^2 y_{13}^2 y_{23}^2\,
 z^2\left[(nx_2)x_{1}^2-(nx_1)x_{2}^2\right]^2
 D_{4433}(x_1,x_2,0,nz)\,,
\end{align}
where $y_{ij}^2$ are defined in \re{Born} with all $Y_{3,AB}$ vanishing except $Y_{3,14}=-Y_{3,41}=1/2$. Here the product
of $y-$variables keeps track of the $R-$charges of the operators whereas the nontrivial dependence on $x-$variables is described by
$D-$function defined in Appendix~\ref{app:B}. 

To extract the correlation function $\vev{O_{\bf 20'}(1) O_{\bf 20'}(2)  O_S(0)}$, we expand
the expression in the second line of \re{OO-nonloc} in powers of $z$ and decompose it over the conformal partial waves.
In this way we find that \re{OO-nonloc} receives a nonvanishing contribution from only one partial wave with $S=2$. In other words, 
the three-point correlation function in the semiclassical approximation is different from zero only for twist-two operators with spin $S=2$:
\begin{align}\notag\label{S=2}
{}&  \vev{O_{\bf 20'}(1) O_{\bf 20'}(2)  O_S(0)}_{\rm inst} 
 \\
{}&  = - \delta_{S,2} 
 {9g^{2}\over 10\pi^2}   \left(  \e^{2\pi i \tau}+ \e^{-2\pi i \bar \tau} \right)
 {y_{12}^2 y_{13}^2 y_{23}^2\over  (4\pi^2)^3x_{12}^2 x_1^2 x_2^2}\left[{2(nx_1)\over x_1^2} -{2(nx_2)\over x_2^2}\right]^2 \,,
\end{align}
where we have added the contribution from the anti-instanton. For $S=0$, this correlation function is protected from quantum corrections.
For higher spin $S>2$, the instanton corrections to \re{S=2} scale as $O(g^4  \e^{2\pi i \tau})$ at least. For $S=2$ we verified 
that the relation \re{S=2} divided by the Born level result coincides with the analogous
expression for the Konishi operator \re{OOK-res}. This is not surprising given the fact that the two operators
$O_{S=2}$ and $K$ belong to the same $\mathcal N=4$ supermultiplet, but rather
serves as a nontrivial check of our calculation.

It is straightforward to extend the above considerations to the two-point correlation function of twist-two operators. Computing the
leading instanton correction to the two-point correlation function of light-ray operators \re{O-LC} and projecting them onto
operators $\mathcal O_S$ with a help of \re{O-LC}, we obtain (see \cite{Belitsky:2005gr} for details on the projection procedure)
\begin{align}\label{OObar}
\vev{\mathcal O_S(x) \bar{\mathcal O}_{S'}(0)}_{\rm inst} = -  \delta_{SS'} \delta_{S,2}  {81\over 5  \epsilon} (x^2)^{-\epsilon} \lr{g^{2}\over 4\pi^2}^2 \lr{ \e^{2\pi i\tau} + \e^{-2\pi i\bar \tau} }
 {[2(xn)]^{2S}\over (4\pi^2)^2  (x^2)^{2+2S}}  \,.
\end{align}
In other words, the instanton corrections vanish for all spins except for $S=2$. In the latter case, they generate the same correction to the scaling 
dimension of the operator $\mathcal O_{S=2}$ as to the Konishi operator \re{anom-K}.
We recall that the two operators belong to the same supermultiplet and their anomalous dimension ought to coincide.

\subsection{Generalization to the $SU(N)$ gauge group}\label{sec:SU(N)}

Having determined the contribution of a single (anti)instanton to correlation functions in $\mathcal N=4$ SYM for the $SU(2)$
gauge group, we can now generalize the above results to the $SU(N)$ gauge group and, in addition, take into account the contribution
of an arbitrary number of (anti)instantons at large $N$.

The instanton for the $SU(N)$ gauge group has $8N$ fermion modes. Among them there are $16$ exact supesymmetric and superconformal 
zero modes, $\xi$ and $\bar\eta$, respectively. The remaining $8N-16$ `nonexact' fermion modes do not correspond to any symmetry and the corresponding $SU(N)$ instanton action $S_{\rm inst}$ develops a nontrivial dependence on these modes. This leads to significant simplification in computing the correlation functions. As in the previous case, for the instanton
correction to be different from zero, all fermion modes should be saturated. Then, in the semiclasscial approximation, the exact modes
are absorbed by the instanton profile of the operators whereas the nonexact modes are saturated by $S_{\rm inst}$. As a consequence,
the contribution of the nonexact modes to the correlation functions factorizes into a univeral $N-$dependent factor \cite{Dorey:1998xe,Dorey:2002ik}
\footnote{Strictly speaking, this relation holds for the so-called minimal correlation functions \cite{Green:2002vf}, for which the product of $n$ operators soaks up the sixteen exact
fermion modes. It is straightforward to verify that the correlation functions \re{S=2} and \re{OObar} are indeed minimal.}
\begin{align}\label{su(n)}
\vev{O(1)\dots O(n)}_{\rm SU(N),\,\rm 1-inst} = \kappa_N \vev{O(1)\dots O(n)}_{\rm SU(2),\, 1-inst} \,,
\end{align}
where $\kappa_N$ takes into account both  the embedding of the
$SU(2)$ instanton in $SU(N)$ and integration over the nonexact modes
\begin{align}
\label{kappan}
\kappa_N = {2^{3-2N}\Gamma(2N-1)\over  \Gamma(N) \Gamma(N-1)} = {2\Gamma(N-1/2)\over \sqrt{\pi} \, \Gamma(N-1)}
\ \stackrel{N\to\infty}{\to} \  {2\sqrt{N}\over \sqrt{\pi}}\,.
\end{align}
In application to the Konishi operator, we can use \re{su(n)} together with \re{anom-K} and \re{c-K} to get its anomalous dimension and OPE
coefficient for the $SU(N)$ gauge group 
\begin{align}\notag\label{res-SU(N)}
{}& \gamma_{K} = - {27 \kappa_N\over 5(N^2-1)}\lr{g^{2}\over 4\pi^2}^2 \lr{ \e^{2\pi i\tau} + \e^{-2\pi i\bar \tau} } \,,
\\
{}& C_{K} = C^{(0)}_{K}\bigg[1- 
 {9  \kappa_N\over 10 (N^2-1)} {g^2\over 4\pi^2} \lr{\e^{2\pi i \tau}+\e^{-2\pi i \bar \tau}} \bigg]\,.
\end{align} 
Here we inserted an additional factor of $3/(N^2-1)$  to account for  $N-$dependence of  two- and
three-point correlation functions in the Born approximation (see Eq.~\re{Born}).

The relations \re{res-SU(N)} can be further generalized to include the contribution of multi-instantons. As was shown in \cite{Dorey:1999pd},
the calculation simplifies dramatically in the large $N$ limit due to the fact that the integration over the moduli space of $n-$instantons is 
dominated by the saddle-point. Repeating the analysis of \cite{Dorey:1999pd}, we find that in this limit the profile of the Konishi operator in the 
$n-$instanton background is proportional to its one-instanton expression. This makes the evaluation of instanton corrections
to the Konishi operator very similar to that performed in  \cite{Dorey:1999pd} for the half-BPS operator. In this way, going through
the calculation we find
the generalisation of \re{res-SU(N)}
\begin{align}\notag\label{res-n-inst}
{}& \gamma_{K}\Big|_{n-\rm inst} = - {54 \over  N^{3/2}}\lr{g^{2}\over 4\pi^2}^2  n^2 (Q_n+ \bar Q_n) \,,
\\
{}& C_{K} \Big|_{n-\rm inst}  = C^{(0)}_{K}\bigg[1- 
 {9 \over 5N^{3/2}} {g^2\over 4\pi^2} n^3 (Q_n+ \bar Q_n)\bigg]\,,
\end{align} 
where we added the contribution of $n$ anti-instantons and introduced
\begin{align}
Q_n= {n^{-7/2}\over \sqrt{\pi}}  \e^{2\pi i n \tau} \sum_{d|n} {1\over d^2}\,,
\end{align}
where the sum runs over the positive divisors of $n$. We would like to emphasize that the relations \re{res-n-inst} hold up to corrections suppressed by powers of $1/N$ 
and $g^2$. The latter come from taking into account quantum fluctuations around the instanton configuration.
  
\section{Instanton corrections to higher spin operators from crossing symmetry} 
\label{bootstrapping}

In the previous section we have computed the instanton corrections to the scaling dimensions of the Konishi and twist-two operators, 
as well as to the OPE coefficients defining their contribution to the product of two half-BPS operators. Using these results we can determine the leading instanton contribution to the four-point correlation function 
\begin{align}\label{G4-def}
G_4=\vev{O_{\bf 20'}(1)\dots O_{\bf 20'}(4)}
\end{align}
at short 
distances, $x_1\to x_2$, and in the light-like limit, $x_{12}^2\to 0$. In this section we use this information, together with crossing symmetry, in order to compute instanton corrections to twist-four operators with large spin. 

\subsection{Properties of the correlation function}  

The four-point correlation function of half-BPS operators in $\mathcal N=4$ SYM with $SU(N)$ gauge group has the following
structure \cite{Eden:2000bk}
\begin{align} \label{G4-gen}
G_4
=  {2(N^2-1) \over (4\pi^2)^4}{y_{12}^4 y_{34}^4\over x_{12}^4 x_{34}^4} \times \big(\mathcal G_{\rm short} +{\mathcal G}_{\rm long}\big)
 \,,
\end{align}
where $\mathcal G_{\rm short}$ and ${\mathcal G}_{\rm long}$ denote the contributions from (semi-)short multiplets and
from long multiplets, respectively. The former contribution does not depend on the coupling constant, whereas the latter 
can be expressed in terms of a single function $\mathcal A_{\rm long}(u,v)$ of the conformal cross-ratios
\begin{equation}
{\mathcal G}_{\rm long}= {
(z-\alpha)(z-\bar\alpha)(\bar z-\alpha)(\bar z-\bar\alpha)\over  (\alpha\bar\alpha)^2}\times 
{\mathcal A_{\rm long}(u,v)\over v^2 }\,.
\end{equation}
The prefactor carries the $R-$charge dependence of the operators and we have introduced the following notation
\begin{align}\label{z}
 u = \frac{x_{12}^2x_{34}^2}{x_{13}^2x_{24}^2} = z\bar z\,,\qqqquad  v=\frac{x_{23}^2x_{14}^2}{x_{13}^2x_{24}^2} =  (1-z)(1-\bar z)
\end{align}
and similar for $\alpha\bar \alpha={y_{12}^2y_{34}^2}/{(y_{13}^2y_{24}^2)}$ and $(1-\alpha)(1-\bar \alpha)={y_{23}^2y_{14}^2}/{(y_{13}^2y_{24}^2)}$. 
The function $\mathcal A_{\rm long} (u,v)$ admits a decomposition in terms of  super-conformal blocks \cite{Dolan:2001tt}
\begin{equation}\label{A-long}
\mathcal A_{\rm long}(u,v) = v^2  \sum_{ S, \, \Delta  }a_{\Delta,S}\, u^{(\Delta-S)/2} g_{\Delta+4,S}(u,v)\,,
\end{equation}
where the sum runs over superconformal primary operators (and hence in the singlet of $SU(4)$) with even Lorentz spin $S$ and scaling dimension $\Delta\ge S+2$  and $a_{\Delta,S}$ is the square of the canonically normalised OPE coefficient.
The contribution from super-conformal descendants is taken into account by the super-conformal blocks, where 
\begin{equation}
\label{gdef}
g_{\Delta,S}(u,v) = \left(-\frac{1}{2} \right)^S \frac{1}{z-\bar z}\left[z^{S+1} k_{\Delta+S}(z) k_{\Delta-S-2} (\bar z) - \bar z^{S+1} k_{\Delta+S}(\bar z) k_{\Delta-S-2}(z)\right],
\end{equation}
with $k_\beta(z) = {}_2F_1(\beta/2,\beta/2,\beta;z)$ and complex $z$ and $\bar z$ variables defined in \re{z}. 

It is convenient to decompose
$\mathcal A_{\rm long} (u,v)$ into the free-theory result $ \mathcal A_{\rm Born}$ 
plus the quantum (coupling dependent) contribution $\mathcal A$
\begin{equation}\label{A-long-1}
\mathcal A_{\rm long} (u,v) = \mathcal A_{\rm Born}(u,v)+ \mathcal A(u,v)\,.
\end{equation}
The explicit expression for $\mathcal A_{\rm Born}(u,v)$ is not needed for our purposes, but it can be derived from the analysis of \cite{Dolan:2004iy}.
At weak coupling, the expansion of $\mathcal A (u,v)$ runs in powers of  't Hooft coupling constant $a=g^2 N/(4\pi^2)$ and
(anti) instanton weight factors, $\e^{2\pi i\tau}$ and $\e^{-2\pi i\bar\tau}$. To leading order in these parameters we have 
\begin{equation}
\label{A0}
\mathcal A (u,v) = -{a\over 4}uv \bar D_{1111}(u,v) +{15 \,\kappa_N\over 2(N^2-1)} \lr{\e^{2\pi i \tau}+\e^{-2\pi i\bar \tau}} u^2v^2 \bar D_{4444}(u,v)
+\dots\,,
\end{equation}
where the $\bar D$-functions are introduced in appendix~\ref{app:B}.
Here the dots denote subleading terms suppressed by powers of the expansion parameters. Higher order perturbative corrections to 
$\mathcal A (u,v)$ were found in \cite{Eden:2011we,Eden:2012tu,Drummond:2013nda}.

Invariance of $G_4$ under the exchange of any pair of points leads to the crossing symmetry relations\footnote{While the free theory contributions $\mathcal G_{\rm short}$ and $\mathcal A_{\rm Born}(u,v)$ mix with each other.}
\begin{equation}
\label{crossing}
\mathcal A (u,v) = \mathcal A (v,u) = v^2 \mathcal A \left({u\over v},{1\over v}\right)\,.
\end{equation}  
Each term on the right-hand side of \re{A0} satisfies this relation.
  
\subsection{Instanton corrections to light-cone asymptotics}  
      
In the light-like limit $x^2_{12}\to 0$, or equivalently $u\to 0$, the leading asymptotic behavior of \re{A-long-1} comes from the contribution
of twist-two operators with scaling dimension $\Delta_S=2+S+\gamma_S$
\begin{align}\label{A-asym}
\mathcal A_{\rm long} (u,v) =u v^2  \sum_{S/2 \in \mathbb Z_+}  a_{\Delta_S,S}\, u^{\gamma_S/2} f_{\Delta_S+4,S}(v) + O(u^2)\,,
\end{align}
where the collinear conformal block $f_{\Delta,S}(v)$ describes the small $u$ limit of \re{gdef}
\begin{align}
f_{\Delta,S}(v)=g_{\Delta,S}(0,v)=\lr{v-1\over 2}^S  {}_2F_1\left( \frac{\Delta+S}{2},\frac{\Delta+S}{2},\Delta+S ; 1-v \right)\,.
\end{align}
The first term on the right-hand side of \re{A-asym} with $S=0$ corresponds
to the Konishi supermultiplet. It gives the leading asymptotic behaviour of $\mathcal A(u,v)$ at short distances, $x_{12}\to 0$,
or equivalently $u\to 0$ and $v\to 1$.  

According to \re{A-long-1} and \re{A0}, the instanton correction to $\mathcal A_{\rm long} (u,v)$ takes the form
\begin{align}\label{A-sub}
\mathcal A_{\rm inst}(u,v) = \mathcal A^{(1)}_{\rm inst} + {g^2\over 4\pi^2} \mathcal A^{(2)}_{\rm inst} + O(g^4)\,,
\end{align}
where $ \mathcal A^{(1)}_{\rm inst}\sim  u^2v^2 \bar D_{4444}$ is given by the second term on the right-hand side of \re{A0} and $\mathcal A^{(2)}_{\rm inst}$ is
the first subleading correction that we shall discuss in a moment. Since $\mathcal A^{(1)}_{\rm inst}(u,v)$ scales as $O(u^2 \ln u)$ in the light-cone limit, 
it does not affect the leading asymptotic behaviour \re{A-asym}. This is in agreement with the known fact that the scaling dimensions and the OPE
coefficients of the Konishi and twist-two operators do not receive instanton corrections at leading order, see {\it e.g.}  \cite{Intriligator:1999ff} and \cite{Arutyunov:2000im} \footnote{Although the claim in  \cite{Intriligator:1999ff} is that this is true in general, their general arguments are only valid in the semiclassical approximation.}.  

In the previous section we have shown that, among all twist-two operators, only those belonging to the Konishi supermultiplet 
receive $O(g^2)$ instanton corrections to their OPE coefficients and $O(g^4)$ corrections to their scaling dimensions. Together
with \re{A-asym} this allows us to fix the small $u$ behaviour of the subleading instanton corrections to \re{A-sub}:
\begin{eqnarray}\notag
\mathcal A^{(2)}_{\rm inst}(u,v) &=&a^{\rm (inst)}_{2,0} u v^2  f_{6,0}(v) + O(u^2)\,, \\[2mm]
\label{smallu}
&=& 30\, a^{\rm (inst)}_{2,0}  \,u v^2  \frac{ 3-3 v^2+\left(v^2+4 v+1\right) \log  v  }{(v-1)^5} + O(u^2)\,,
\end{eqnarray}
where we neglected $O(g^4)$ corrections to the scaling dimension $\Delta_{S=0} = 2 +\gamma_K$, since they do not enter at this order.  Here $a^{\rm (inst)}_{2,0}$ denotes the instanton correction to the OPE coefficient of the Konishi operator. It can be written in terms of the free-theory coefficient\footnote{The OPE coefficient for the Konishi operator in the free theory is $a^{(0)}_{2,0}=1/6$, but we will not need its explicit value in this section.}  
\begin{align}\label{a-inst}
a^{\rm (inst)}_{2,0}= - a^{(0)}_{2,0} {9\kappa_N\over 5(N^2-1)}\lr{\e^{2\pi i \tau}+\e^{-2\pi i \bar \tau}}\,.
\end{align}
The ratio $a^{\rm (inst)}_{2,0}/a^{(0)}_{2,0}$ coincides with the instanton correction to $(C_K/C_K^{(0)})^2$ (see Eq.~\re{res-SU(N)}).
Note that $\mathcal A^{(2)}_{\rm inst}(u,v)$ does not contain $O(u \log u)$ terms. This should of course be the case, as there are no instanton corrections to the anomalous dimensions of twist-two operators at this order. Such corrections first appear at order $O(g^4)$.
In the next subsection we use crossing symmetry of $\mathcal A_{\rm inst}(u,v)$ together with the small $u$ behaviour (\ref{smallu}) in order to compute instanton corrections to certain 
higher spin operators.  
 
\subsection{Crossing symmetry and higher spin operators}

Before proceeding, let us make an important comment. As follows from the light-cone asymptotic behaviour of 
$\mathcal A^{(1)}_{\rm inst}(u,v) \sim u^2 \log u$, the anomalous dimension of operators of twist four and higher do receive 
instanton corrections at the leading $O(g^0)$ order \cite{Arutyunov:2000im}. The conformal partial wave analysis shows
\cite{Alday:2014tsa} that only operators with spin zero receive such corrections. We show below that the situation is very 
different at order $O(g^2)$.  

Let us examine $\mathcal A_{\rm inst}(u,v)$ in the double light-cone limit $u,v\to 0$. Combining the leading asymptotics (\ref{smallu}) with the crossing relation $\mathcal A_{\rm inst}(u,v)=\mathcal A_{\rm inst}(v,u)$ we infer that in the small $u,v$ limit $\mathcal A_{\rm inst}(u,v)$ should contain the following term 
\begin{eqnarray}
\mathcal A_{\rm inst}(u,v) &=&-30 {g^2\over 4\pi^2} a^{\rm (inst)}_{2,0}  v \, u^2 \log u + \cdots\,.
\end{eqnarray}
Following \cite{Komargodski:2012ek,Fitzpatrick:2012yx} we then try to answer how to get such asymptotics from a conformal partial wave  expansion. The crucial observation is that, due to the presence of $u^2 \log u$ term, this must come from twist-four operators with   anomalous dimension $\gamma_{4,S}=O(g^2 \lr{\e^{2\pi i \tau}+\e^{-2\pi i \bar \tau}})$. More precisely, the expansion (\ref{A-long}) should contain a contribution from a tower of 
twist four operators with Lorentz spin $S$ such that
\begin{align}\label{crossingrel}
 v^2  \sum_{S=0,2,\cdots}  a_{4,S}\, u^2 u^{\gamma_{4,S}/2} f_{8+S,S}(v)\big|_{u^2 \log u} = -30  {g^2\over 4\pi^2}  a^{\rm (inst)}_{2,0} \, v u^2 \log u + {\cal O}(v^2)\,,
\end{align}
or equivalently
\begin{align}\label{sumrule}
 \frac{1}{2}   \sum_{S=0,2,\cdots}  a_{4,S}\, \gamma_{4,S} f_{8+S,S}(v)\Big|_{1/v} = -\frac{30}{v}  {g^2\over 4\pi^2} a^{\rm (inst)}_{2,0}  \,.
\end{align}
It is important to emphasise that, as opposed to twist-two operators, for a given spin $S$ the sum on the left-hand side receives the
contribution from many twist-four operators. 
To simplify formulae we do not add an additional index to distinguish such operators.

A very important point about \re{sumrule} is that, given that each term on the left-hand side diverges only logarithmically as $v\to 0$, we need an infinite number of them in order to reproduce the power law divergence $1/v$ on the right-hand side of \re{sumrule}. Furthermore, the divergence will come from the region with large spin, $S\gg 1$. The corresponding OPE coefficients $a_{4,S}$ in the free theory were found in \cite{Dolan:2004iy}. In the large spin limit they reduce to 
\begin{equation}
a_{4,S} = a^{(0)}_{2,0} \frac{\sqrt{\pi}  (N^2-1) S^{5/2}}{ 2^{S+8}} + O(g^2)\,.
\end{equation}
The leading asymptotic behaviour of the left-hand side of (\ref{sumrule}) for $v\to 0$ can be computed following \cite{Alday:2010zy} (see footnote 19 there). Matching the leading $1/v$ terms on both sides of (\ref{sumrule}) we obtain
\begin{equation}
\gamma_{4,S} = - \frac{\alpha}{S^2}\,,\qqqquad 
\alpha=\frac{480}{N^2-1} \frac{g^2}{4\pi^2} { a^{\rm (inst)}_{2,0} \over a^{(0)}_{2,0} }\,.
\end{equation}
Replacing $a^{\rm (inst)}_{2,0}$ with \re{a-inst}, we arrive at our final expression for the large spin behaviour of the
anomalous dimension of twist-four operators
 
\begin{equation}\label{g4}
\gamma_{4,S} = \frac{864\, \kappa_N}{(N^2-1)^2} \frac{g^2}{4\pi^2}(\e^{2\pi i \tau}+ \e^{-2\pi i \bar \tau}) \frac{1}{S^2}\,.
\end{equation}
We remind that since twist-four operators are degenerate, this anomalous dimension should in principle be understood as an average  weighted 
by the tree-level OPE coefficients. 

The following comments are in order. Firstly, the large spin asymptotics $\gamma_{4,S} \sim 1/S^2$ is consistent with the expected behaviour of anomalous dimensions of double trace operators \cite{Alday:2007mf}. Moreover, $\gamma_{4,S}$ is suppressed
by the factor of $(N^2-1)$ as compared with the anomalous dimension of the Konishi operator (see Eq.~\re{res-SU(N)}) which is
also a characteristic feature of double trace operators.
Secondly, while at order $O(g^0)$ only twist four operators with zero spin receive instanton corrections, at order $O(g^2)$ operators with arbitrarily high spin do. It would be very interesting to compute these corrections 
directly.  Finally, the relation \re{g4} describes the contribution of one-(anti)instanton in the $SU(N)$ gauge group.  Making use
of \re{res-n-inst}, it is straightforward to generalize it to multi-instantons in the large $N$ limit.

\section{Conclusions}

In the present paper we have computed, in the semi-classical approximation, instanton corrections to various correlation functions, involving the half-BPS operator $O_{\bf 20'}$ and the Konishi operator $K$. Our main results are the explicit expressions \re{res-SU(N)} and \re{res-n-inst}
for the leading instanton contribution to the anomalous dimension of the Konishi operator, as well as for the OPE coefficient of the Konishi operator with two half-BPS operators.

In addition, we considered twist-two operators of general spin $S$, and showed that the only operators that receive the leading instanton corrections are those which carry spin $S=2$ and belong to the Konishi supermultiplet. 
Using this information, we derived the asymptotic light-cone behaviour of the correlation function of four half-BPS operators and, then,  
employed the crossing symmetry to determine the instanton contribution to the anomalous dimension of twist-four operators, in the limit of large spin. 

Our computations show a very interesting interplay between semi-classical vs quantum instanton corrections and the symmetries of ${\cal N}=4$ SYM.  An instance of this interplay arises when comparing instanton corrections to the scaling dimensions of Konishi operator
$K$ and its dimension-four supersymmetric descendant $K' \sim \delta_Q^2 \delta_{\bar Q}^2 K$. Superconformal symmetry implies that these two operators should have the same anomalous dimensions. On the other hand, while the leading non-vanishing instanton correction  to $\Delta_K$ comes from a semi-classical computation, from the perspective of $\Delta_{K'}$ computing the same correction  would require
going through a highly nontrivial analysis of quantum fluctuations!  Something similar happens when considering the 
four-point correlation function of half-BPS operators. Subleading instanton corrections to this correlator involve including quantum fluctuations, but on the other hand, the asymptotic behaviour of such corrections at short distances and on the light-cone is controlled through the OPE by two- and three-point correlation functions, in which instanton corrections at the same order in the coupling constant can be derived from a semi-classical computation. All this seems to hint at existence of some hidden structure underlying instanton corrections in  ${\cal N}=4$ SYM.

There are several directions in which this work can be extended. As was mentioned in the Introduction, the spectrum of the dilatation 
operator in $\mathcal N=4$ SYM should be invariant under the $S-$duality. Viewed as functions of the coupling constant, the scaling 
dimensions of operators carrying the same quantum numbers with respect to global symmetries cannot cross each other and 
they should be invariant under modular transformations. Since modular invariant functions independent on $\theta-$angle
ought to be constant, the scaling dimensions should have a nontrivial dependence on $\theta$. One of the direct consequences of our 
study is that it opens up the possibility to construct such functions in $\mathcal N=4$ SYM by taking into account the instanton 
corrections.  Our results represent the first explicit calculation of the instanton correction to Konishi operator which is 
the lowest unprotected operator at weak coupling. The obtained expression \re{res-SU(N)} can be thought of as representing the first 
term in the expansion of the modular invariant function $\Delta_K(\tau,\bar\tau)$ at weak coupling. It would be interesting to try to 
determine the modular properties of $\Delta_K$ for an arbitrary coupling. As a first step in this direction, one can use the
results of this paper to improve an interpolating procedure proposed in \cite{Beem:2013hha,Alday:2013bha}.

Another interesting question concerns $S-$duality properties of the OPE coefficients. In distinction with the 
scaling dimensions, they do not have to satisfy von Neumann--Wigner non-crossing rule \cite{Korchemsky:2015cyx} and, as a consequence, they may
transform nontrivially under the modular transformations of the coupling constant. In general, the properties of structure constants under S-duality are poorly understood. 
Our results may provide the first hints in this direction, for the simplest case, corresponding to the OPE coefficient of two half-BPS operators and a unprotected operator. For the 
Konishi operator the leading instanton correction is given by \re{res-SU(N)}, finding higher order corrections is an open problem.

It would also be interesting to understand better the interplay between semi-classical and quantum instanton corrections mentioned
above. Our results seem to hint at unexpected simplifications when considering quantum corrections around instanton backgrounds in ${\cal N}=4$ SYM.  In addition to this, we demonstrated that the semiclassical result for
the Konishi operator combined with the crossing symmetry leads to a definite prediction for the instanton correction to the
scaling dimensions of twist-four operators with large spin. Computing this correction directly would require to include a quantum 
fluctuations. This remains a largely unexplored subject.
 
\section*{Acknowledgments}

We are grateful to Agnese Bissi, Nick Dorey, Jaume Gomis, Stefano Kovacs and Pierre Vanhove for useful discussions. L.F.A. and G.P.K. acknowledge Nordita and Perimeter Institute,
respectively, for hospitality where part of this work has been done. The work of L.F.A. was supported by ERC STG grant 306260. L.F.A. is a Wolfson Royal Society Research Merit Award holder.  This work of G.P.K. was supported in part by
the French National Agency for Research (ANR) under contract StrongInt (BLANC-SIMI-
4-2011).

\appendix
   
\section{Conventions}

In this appendix we summarize our conventions for dealing with spinor indices. For an arbitrary four-dimensional vector $x^\mu$
we associate a $2\times 2$ matrix
\begin{align}
x_{\alpha\dot\beta} = x_\mu (\sigma^\mu)_{\alpha\dot\beta} \,,
\end{align}
where $\sigma^\mu=(1,\boldsymbol \sigma)$ is defined in terms of Pauli matrices. The indices are raised and lowered with a help of
the completely antisymmetric tensor  
\begin{align}
x^\alpha_{\dot\beta} =\epsilon^{\alpha\beta}x_{\beta\dot\beta} \,,\qqquad
x_{\alpha}^{\dot\beta} =x_{\alpha\dot\alpha}\epsilon^{\dot\alpha\dot\beta} \,,\qqquad
x^{\dot\alpha\beta} = \epsilon^{\beta\alpha}x_{\alpha\dot\beta} 
\epsilon^{\dot\beta\dot\alpha}\,,
\end{align}
with $\epsilon_{\alpha\beta} \epsilon^{\alpha\gamma} = \delta_\beta^\gamma$ and  
$\epsilon_{\dot\alpha\dot\beta} \epsilon^{\dot\alpha\dot\gamma} = \delta_{\dot\beta}^{\dot\gamma}$. Using these definitions we find
\begin{align}\notag
{}& x^2  
=\frac12 x_{\alpha\dot\alpha}x_{\beta\dot\beta}\epsilon^{\alpha\beta}\epsilon^{\dot\alpha\dot\beta}
\,,\qqqquad
 x^{  \beta}_{\dot\beta} x^{\dot\alpha}_{\beta} = -x^{\dot\alpha \beta} x_{\beta \dot\beta} =  x^2 \delta_{\dot\beta}^{\dot\alpha}\,.
\end{align}  
For derivatives we have in the similar manner
\begin{align} 
{}& \partial_{\alpha\dot\beta} = \partial_\mu (\sigma^\mu)_{\alpha\dot\beta} \,,&&
\partial_{\alpha\dot\beta}\partial^{\dot\beta\beta} = - \delta_\alpha^\beta \,\Box
 \,,&&
 \partial^\beta_{\dot\beta}\partial_\alpha^{\dot\beta} = \delta_\alpha^\beta \,\Box\,.
\end{align} 
The gauge field strength tensor $F_{\mu\nu}$ can be decomposed into (anti) self-dual part as
\begin{align}
F_{\alpha\beta} = i F_{\mu\nu} (\sigma^\mu)_{\alpha\dot\alpha} (\sigma^\nu)^{\dot\alpha}_\beta\,,\qqqquad
F_{\dot\alpha\dot\beta} = i F_{\mu\nu} (\sigma^\mu)^\alpha_{\dot\alpha} (\sigma^\nu)_{\alpha \dot \beta}\,,
\end{align}
with $ (\sigma^\nu)^{\dot\alpha}_\beta= (\sigma^\nu) _{\beta\dot\beta}\epsilon^{\dot\beta\dot\alpha}$ and
$(\sigma^\mu)^\alpha_{\dot\alpha}=\epsilon^{\alpha\beta}(\sigma^\mu)_{\beta\dot\alpha}$.

\section{Definition of the $D-$functions}\label{app:B}

The $D-$functions are defined as
\begin{align}
D_{\Delta_1\Delta_2\Delta_3\Delta_4}(x_1,x_2,x_3,x_4) = \int d^4 x_0 \int {d\rho\over\rho^5} \prod_{i=1}  \left( {\rho \over \rho^2 + x_{i0}^2} \right)^{\Delta_i}\,,
\end{align}
where $x_{i0} = x_i-x_0$.
Using Schwinger parameterization together with the Symanzik star formula we get
\begin{align}
D_{\Delta_1\Delta_2\Delta_3\Delta_4} {}& =2K
\int_0^\infty \prod_i dt_i \, t_i^{\Delta_i-1} \e^{-\sum_{i<j} t_i t_j x_{ij}^2 }
=K \int \prod_{i<j}^4 {d \delta_{ij}\over 2\pi i} \Gamma(-\delta_{ij}) (x_{ij}^2)^{\delta_{ij}}\,,
\end{align} 
where $K=\pi^2  {\Gamma\lr{\frac12 \sum \Delta_i-2} / (2\prod_i \Gamma(\Delta_i))}$ and the integration in the last relation goes parallel to the imaginary axis. The variables $\delta_{ij}$ are not independent and satisfy 
the relations
\begin{align}
\delta_{ij} = \delta_{ji}\,,\qqqquad \sum_{j\neq i}^4 \delta_{ij} = - \Delta_i\,,
\end{align}
leaving only two variables independent. It is convenient to choose the latter as $\delta_{12} =  j_1$ and $\delta_{23} =  j_2$.
In this way, we obtain
\begin{align}
D_{\Delta_1\Delta_2\Delta_3\Delta_4} = {K\over (x_{13}^2)^{\Delta_1}(x_{24}^2)^{\Delta_2}}\lr{x_{13}^2 x_{34}^2\over x_{14}^2}^{\frac{\Delta_1-\Delta_3}2}\lr{x_{14}^2 x_{34}^2\over x_{13}^2}^{\frac{\Delta_2-\Delta_4}2} \bar D_{\Delta_1\Delta_2\Delta_3\Delta_4} 
(u,v)\,,
\end{align}
where the $\bar D-$function only depends on conformal cross-ratios \re{z}
\begin{align}\notag
\bar D_{\Delta_1\Delta_2\Delta_3\Delta_4} {}& =\int_{-i\infty}^{i\infty}{ dj_1 dj_2\over (2\pi i)^2} u^{j_1} v^{j_2}
\Gamma(
j_1+j_2+\Delta_2)
\Gamma( j_1+j_2 +\Delta-\Delta_4)
\\
{}& \times  \Gamma(-j_1)\Gamma(-j_2)\Gamma\left(-j_2 +\Delta-\Delta_2-\Delta_3\right)\Gamma\left( - j_1- \Delta+\Delta_3+\Delta_4\right)\,,
\end{align}  
with $\Delta=\sum_{i=1}^4 \Delta_i/2$. Here the integration contours are chosen in such a way that the
poles generated by the product of gamma-functions in the first and second lines are located on the different sides.

\bibliographystyle{JHEP} 


\providecommand{\href}[2]{#2}\begingroup\raggedright\endgroup

\end{document}